\documentclass[%
 reprint,
 amsmath,amssymb,
 aps,
pra,
]{revtex4-2}

\usepackage{graphicx}
\usepackage{subfigure}
\usepackage{dcolumn}
\usepackage{bm}
\usepackage[utf8]{inputenc}
\usepackage{braket}
\usepackage{xcolor}
\usepackage{microtype}
\usepackage{siunitx}
\usepackage{physics}
\usepackage{siunitx}
\usepackage{hyperref}

\usepackage{tikz}
\usetikzlibrary{external}
\usepackage{pgfplots}
\pgfplotsset{compat=1.7}
\AtBeginDocument{\RenewCommandCopy\qty\SI} 
\definecolor{color1}{HTML}{648FFF} 
\definecolor{color2}{HTML}{785EF0} 
\definecolor{color3}{HTML}{DC267F} 
\definecolor{color4}{HTML}{FE6100} 

\definecolor{gray1}{HTML}{111111} 
\definecolor{gray2}{HTML}{555555} 
\definecolor{gray3}{HTML}{999999} 

\pgfplotsset{compat=1.11}

\usepackage{changes}
\definechangesauthor[name={Giovanni Cerchiari}, color=purple]{GC}
\definechangesauthor[name={Lorenz Panzl}, color=orange]{LP}
\definechangesauthor[name={Yannick Weiser}, color=green]{YW}
\definechangesauthor[name={Tommaso Faorlin}, color=color1]{TF}
\definechangesauthor[name={Thomas Monz}, color=color3]{TM}
\definechangesauthor[name={Rainer Blatt}, color=teal]{RB}

\begin{document}

\preprint{APS/123-QED}

\title{Controlling the spontaneous emission of trapped ions}

\author{Tommaso Faorlin}
\author{Yannick Weiser}%
\author{Lorenz Panzl}
\author{Thomas Lafenthaler}
\author{Rainer Blatt}
\altaffiliation[Also at ]{Institut fur Quantenoptik und Quanteninformation, ¨Osterreichische¨
Akademie der Wissenschaften, Technikerstraße 21a, 6020 Innsbruck, Austria.}
\author{Thomas Monz}
\altaffiliation[Also at ]{AQT, Technikerstraße 17, 6020 Innsbruck, Austria.}
\author{Giovanni Cerchiari}
\altaffiliation[Also at ]{Naturwissenschaftlich-Technische Fakultät, Universität Siegen, Walter-Flex-Straße 3, 57068 Siegen, Germany.}
\email{Second.Author@institution.edu}
\affiliation{Universit\"at Innsbruck, Institut f\"ur Experimentalphysik, Technikerstraße~25a, 6020~Innsbruck, Austria}
\author{Benjamin Yadin}
\author{Stefan Nimmrichter}
\affiliation{Naturwissenschaftlich-Technische Fakultät, Universität Siegen, Walter-Flex-Straße 3, 57068 Siegen, Germany.}%
\author{Gabriel Araneda}
\affiliation{Department of Physics, University of Oxford, Clarendon Laboratory, Parks Road, Oxford OX1 3PU, United Kingdom.}%

\date{\today}

\begin{abstract}
We propose an experimental setup for manipulating the spontaneous emission of trapped ions, based on a spatial light modulator. Anticipated novelties include the potential to entangle more than two ions through a single photon detection event and control the visibility for spatially distinguishable emitters. The setup can be adapted to most of the existing ion traps commonly used in quantum technology.
\end{abstract}
\maketitle


\section{\label{sec:intro}Introduction}
The process of spontaneous emission (SE) of radiation is a phenomenon in which excited states in matter release energy into the environment as electromagnetic radiation~\cite{Milonni1984}. In trapped atoms, this process is crucial for the realization of Doppler cooling and quantum-state detection \cite{leibfried2003quantum}. In addition, the exchange of spontaneously emitted photons is an alternative for entanglement generation to standard methods based on motional coupling of ions that interact at close proximity within the same trapping volume~\cite{SorensenMolmer99}. For example, SE gives experimenters the possibility to distribute quantum resources beyond the trapping device in quantum networks~\cite{moehring2007entanglement,Stephenson2020} for remote  sensing, and precision geodesy~\cite{Beloy2021}. Photon-mediated entanglement is appealing for quantum computing also within the same Quantum Processing Unit (QPU) because photon transfer might prove faster than transporting the ion with the associated information (shuttling). Ion shuttling is one of the most time-costly overheads of modern quantum computing chip architectures~\cite{Fuerst2014, Hilder2022} and the advancing of a new general operational framework utilizing the `photon bus' as opposed to the conventional `motion bus', holds significant potential to expedite computations reliant on entanglement distribution~\cite{Buhrman2003, caleffi2022distributed}. Furthermore, high-fidelity entanglement generation is of paramount importance in modular quantum computing architectures, because it allows quantum teleportation of states or gates between the different parts. For local photon-mediated entanglement, primarily two methods have been considered. One approach consists of arranging ions into suitable geometrical configurations with respect to a detector~\cite{Obsil19, Wolf20}, which requires the physical movement of the ions or the detector to control the interaction. In the second approach, a mirror can be utilized to reflect the light emitted by one ion onto another one~\cite{Araneda18}, which eliminates the need to physically rearrange the ions in the trap to entangle them. The presence of a single mirror in the latter approach, limits the entangling protocol to a single pair of ions at a time. In this article, we propose to overcome this limitation by using a phase Spatial Light Modulator (SLM).

The setup can also be used to control the phase front of light emitted by multiple ions. This control enables interference of the ions' spontaneous emission while the ions remain spatially distinguishable, presenting a novel experimental condition that complements parallel studies where interference was achieved under conditions that made the ions indistinguishable~\cite{Obsil19}. Our method combines the studies on light scattering of single ions in half-cavities~\cite{Eschner01} with explorations in multi-ion light interference~\cite{Wolf20, Mährlein20, Skornia01}.  The SLM is adapted to retro-reflect the wave fronts of multi-ion emission effectively providing control over their collective emission while they remain distinguishable in space. 

The paper is organized as follows: 
In Sec.~\ref{sec:setup} a model describing how the emission of the ions can be influenced by the SLM is presented. 
Then, in Sec.~\ref{sec:quantum-operations}, we present how the SLM can be used to program complex and flexible entanglement operations involving more than two ions. Finally, in Sec.~\ref{sec:collective-spontaneous-emission} we describe how the SLM can regulate the spontaneous emission of multiple ions.

\section{\label{sec:setup}Model}

\begin{figure}[ht!]
    \centering
    \includegraphics[scale=0.44]{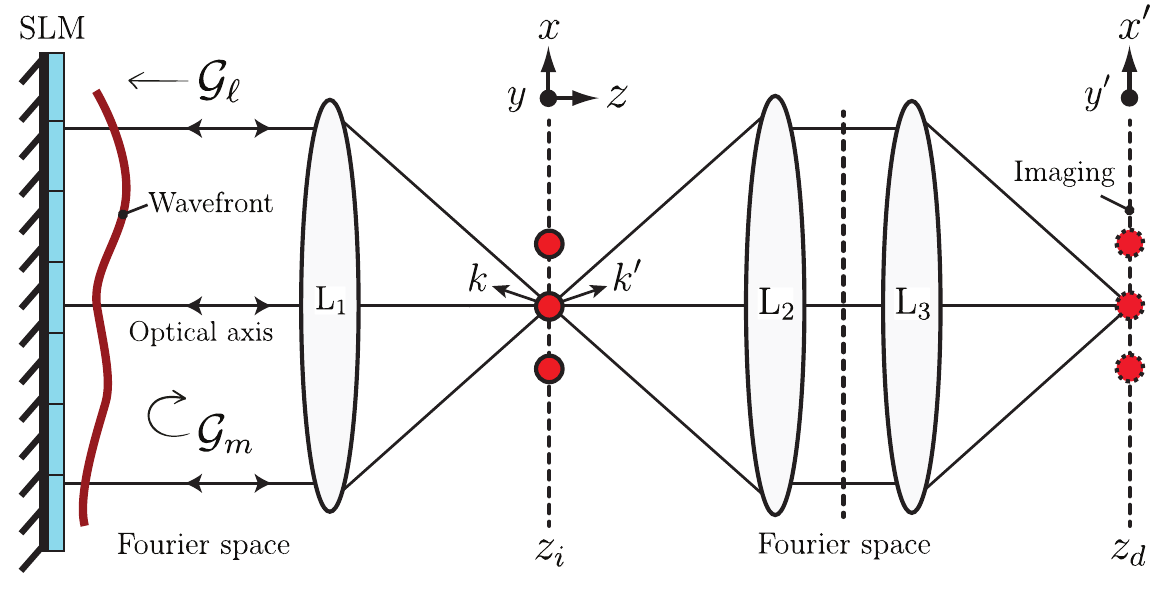}
    \caption{Schematic optical setup. Ions are trapped in the focal plane $(x,y,z=z_i)$ of a confocal lens setup. On the left side in the figure, a SLM modifies the wavefront of the light field emitted by the ions upon back reflection. The reflected field returns towards the ions and interferes with the primary field emitted by the ions in the direction of detection. The SLM can be programmed to obtain a different phase delay of the reflected field at different wave vectors $\bm{k}$. The distance of the lens $L_2$ can be adjusted to collimate a spherical wavefront into plane waves and, with $L_3$, to image the ions onto a detector plane ($x',y',z=z_d$). Collimation into plane waves forms the Fourier transform of the field.}
    \label{fig:optical_setup}
\end{figure}
A scheme depicting the optical setup is presented in Fig.~\ref{fig:optical_setup}. The classical model introduced in this section, aims at explaining how ions are coupled to a boundary condition, and how by programming the latter some applications can be realized. We consider as dipolar emitters $N$ ions located in the focal plane of a confocal pair of lenses. The electric field of the emitted photons of such an arrangement can be modeled by a function $f(\bm{r})$, where $\bm{r}\in(x,y,z=z_i)$ and $z_i$ marks the plane where the ions are. We assume the lenses to be located in the far-field of the ions and simplify the description of the system by applying the paraxial approximation. Under these conditions, the lenses ($L_1$ and $L_2$) map a field distribution $f(\bm{r})$~\cite{goodman2005} located in their focal plane into its spatial Fourier transform (FT) $\tilde{f}\left(\bm{k}\right)$ via 
\begin{equation}
 \label{eq:free_space_paraxial}
     \tilde{f}(\bm{k})\propto\int_{\mathbb{R}^2} d^2\bm{r}\,\,\mathcal{G}_\ell(\bm{k},\bm{r})f(\bm{r}) \; ,
 \end{equation}
where we called $d^2\bm{r}=dxdy$ and where the Green's function $\mathcal{G}_\ell$ is
\begin{equation}
    \label{eq:g_func_paraxial}
    \mathcal{G}_\ell(\bm{k},\bm{r})=e^{-i\bm{k}\cdot\bm{r}}.
\end{equation}
In the last equation, the wave vector $\bm{k}$ parameterizes the $(x,y)$ plane in the far-field~\cite{goodman2005}.

On one side of the confocal setup, a reflective phase SLM is used to send the radiation back to the emitters. This device consists of a matrix of pixels, where each pixel can change the local phase of the impinging wavefront upon back-reflection. This arrangement can be modeled by introducing the Green's function $\mathcal{G}_m$ for the returning field right after reflection. The expression of $\mathcal{G}_m$ is
\begin{equation}
    \mathcal{G}_m(\bm{k},\bm{r}) = \rho\,\tilde{m}(\bm{k}) \mathcal{G}_\ell(\bm{k},\bm{r})\;,
\end{equation}
where $\rho=\epsilon_1\epsilon_2$ accounts for the non-unitary reflection coefficient $\epsilon_1$ of the SLM and the optical losses $\epsilon_2$ through the imaging system. $\tilde{m}(\bm{k})$ is the phase function programmed on the SLM's pixels. The refocusing of the returning light field by the lens $L_1$ performs an additional FT, resulting in the field
\begin{equation}\label{eq:SLM_only}
f_m(\bm{r})=\rho e^{i\psi}(f\ast m)(-\bm{r}) \,,
\end{equation}
where $\ast$ is the convolution of the original function $f$ with $m(\bm{r})$ being the FT of $\tilde{m}(\bm{k})$. The constant phase term $e^{i\psi}$ accounts for the optical path of the light traveling to and from the modulating device. In a real setup, this phase term can be adjusted by controlling the distance of the SLM to the lens with a piezomechanical actuator, or by adding a global bias phase to the programmed phase function. In Eq.~\eqref{eq:SLM_only}, the minus sign on the variable $\bm{r}$ indicates that the retro-reflected light forms an image at the opposite side with respect to the focal point of $L_1$.
The combined system composed by ions and mirror can be considered as a half-cavity~\cite{Eschner01}. When the ions are positioned close to the waist of the half-cavity,  the emitted light is redirected onto the ions themselves and their SE is affected~\cite{Dorner02,Cerchiari2021}. Due to field interference, the total field emerging from the ion plane propagating towards the lens $L_2$ is given as~\cite{cerchiari2021two}
\begin{equation}
\label{eq:field_chain_summed}
    u(\bm{r})=f(\bm{r})+f_m(\bm{r})\,.
\end{equation}
This formula is consistent with the case of a flat mirror ($\tilde{m}(\bm{k})=1$), which forms a reflected image of the emitters dephased according to the round-trip optical path of the reflected light. Any subsequent collimation or refocusing adds additional FTs. For example, if the second lens $L_2$ is used to collimate the spherical waves emitted from the ions' plane into plane waves the light field after the lens can be written as
\begin{equation}
\label{eq:field_detector_theta}
    \tilde{u}(\bm{k}')=\tilde{f}(\bm{k}')+\rho e^{i\psi}\tilde{m}(-\bm{k}')\tilde{f}(-\bm{k}').
\end{equation}
Instead, if the lens is used to image the ion plane onto a detector (equivalent to considering $L_3$ in the complete setup), such as an EMCCD camera, the final field on the detector is $d(\bm{r}')=u(-\bm{r}')$. We report a more detailed derivation of the formulas presented in this section in Appendix~\ref{sec:appendix-detailed-model}, and proceed to illustrate possible applications of this model in the next paragraphs.  

\section{\label{sec:simulations}Applications}
We present two applications for the introduced setup, simulating real experimental conditions and a commercially available SLM (Hamamatsu X15213-16). The code is available online~\cite{cerchiari_2025_14670747} in the form of Jupyter notebooks and Python scripts.  The first application is about the programmable creation of photon-mediated entanglement, using the Cabrillo scheme~\cite{Cabrillo1999}. Here, the SLM acts as a flexible mirror that creates multiple images of the ion chain on the detector plane. In this way, it is possible to superimpose the images of several ions at the plane of detection and generate entangled states by photon detection. The second application is the controlling of the spontaneous emission of a chain of ions. Up to now, control was only demonstrated for single ions~\cite{Eschner01} and in multiple ion systems restricted to configurations that made ions indistinguishable~\cite{Obsil19, Slodička13, Araneda18}. Unlike a mirror, the fact that the SLM is programmable enables experimenters to control non-uniform wave fronts emitted by multiple ions permitting SE control also for ions that remain spatially distinguishable in the imaging plane. We simulate the fields described in Eqs.~\eqref{eq:SLM_only}, \eqref{eq:field_chain_summed}, and \eqref{eq:field_detector_theta} and show which phase masks need to be programmed on the SLM to achieve programmable entanglement and modulated spontaneous emission. 

\subsection{\label{sec:quantum-operations}Remote entanglement}
\begin{figure}[t!]
    \centering
 \includegraphics[scale=0.9]{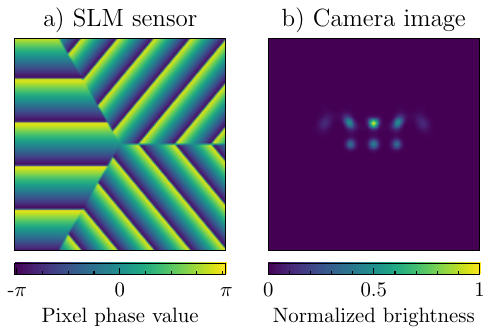}
    \caption{If three ions are entangled, the phase mask programmed on the modulator’s pixels resembles the one shown on the left. Each angular sector of the mask creates an image of the entire ion chain on the detector plane, as illustrated on the right. These images are shifted relative to the original chain by an amount proportional to the frequency and orientation of the pattern (a blazed grating) encoded in the pixels of each sector. The shifts are programmed to ensure that the images of all three distinct ions precisely overlap at the same point on the detector. The single photon that signals entanglement is detected at the brightest spot in (b), where all the ions’ images overlap via the SLM.}
    \label{fig:SLM_entanglement_phasemask}
\end{figure}
Entanglement of two ions by measurement of a single photon can be achieved by implementing the Cabrillo scheme~\cite{Cabrillo1999} with a distant mirror~\cite{Slodička13, Araneda18}. The protocol for two ions relies on the mirror for superimposing the spatial modes of the ions onto a single photon detector~\cite{Araneda18}: the source of an eventual detection of a single photon is therefore indistinguishable. We show how this protocol can be extended to several ions by using the SLM instead of a mirror. We then motivate the phase function to be programmed first and present the details of the generated ions' quantum state.

A reflective SLM can be employed to adjust the direction of a reflected light beam by programming a blazed grating function. This function is implemented as a linear increase of the phase across the SLM, which is equivalent to a tilt of an ordinary mirror, with periodical phase wrapping to keep the phase values within the allowed programmable range.
Following this idea, we can divide the SLM sensor into as many sectors as there are ions to be entangled, in order to create several images of the ion chain and simultaneously superimpose them. In Fig.~\ref{fig:SLM_entanglement_phasemask} a) the three sectors are recognizable by the different pitches and inclinations of the blazed gratings. The sectors should preferably have equal area and divide the SLM symmetrically around the optical axis (as a pie) to balance the reflected power, considering the transfer function of the optical apparatus. In a generic sector we encode a shift operator $\tilde{m}$ that is programmed to displace the image of an ion labeled $j$ onto the detector located at $\bm{r}_d$. The function to be programmed on the sector is
\begin{equation}\label{eq:entanglement_mask}
    \tilde{m}\left(\bm{k}\right) = \exp{i(\phi+\bm{k}\cdot(\bm{r}_j-\bm{r}_d))}\;,
\end{equation}
where $\phi$ is a constant phase shift that can be programmed on the entire sector and $\bm{r}_j$ is the location of the ion on the plane. Figure~\ref{fig:SLM_entanglement_phasemask} depicts one of such configurations to generate three superimposed images of three ions. In some areas of the image plane (brightest point in Figure~\ref{fig:SLM_entanglement_phasemask} b), the photons emitted by the three ions are indistinguishable, which hence can be used to generate entanglement via the Cabrillo scheme. The sectors are re-programmable and re-configurable allowing for a variety of flexible operations.

The entangling protocol can be implemented via detection of spontaneously emitted photons from a cyclic transition. In a generic scenario, we consider $N$ ions, each of them initialized in an unbalanced superposition of two ground states $\ket{g_\pm}$, i.e., $\sqrt{1-p}\ket{g_-} + \sqrt{p}\ket{g_+}$. A global $\pi$-pulse transfers the population of $\ket{g_+}$ to a fast-decaying excited state $\ket{e}$. After spontaneous emission, the state $\ket{e}$ decays back to $\ket{g_+}$, emitting a single photon. In the example of $^{138}$Ba$^+$, $\ket{g_\pm}=\ket{6S_{1/2},m_j=\pm1/2}$ and the $\pi$ pulse can be driven with a \qty{455}{\nano\meter} laser to $\ket{e}=\ket{5P_{3/2},m_j=+3/2}$ so that the decay back to $\ket{g_-}$ is dipole forbidden and logical errors are thus nonexistent. If $p$ is small, the probability of multiple photon events is negligible. If the light emission modes of the different ions are superimposed on a detector, the detection of a single photon event does not provide information about which particular ion emitted the photon, projecting the ions into an entangled state. In case the state $\ket{e}$ has other decay channels, the produced photons can be filtered out using polarization- and wavelength-sensitive photon detectors.

Using the SLM arrangement and the phase mask shown in Fig.~\ref{fig:SLM_entanglement_phasemask}, the collective emission of a single photon can be detected in the image plane, at the brightest spot in Fig.~\ref{fig:SLM_entanglement_phasemask}, which provides indistinguishability of the possible emitters. The quantum state after emission, including electronic levels in tensor product with number states $\ket{n}$ of radiation quanta, is as follows
\begin{equation}
\ket{\psi}=\sum_{n=0}^Np^{n/2}(1-p)^{(N-n)/2}\ket{W_{n,N-n}}\ket{n} \, ,
\end{equation}
where $\ket{W_{n,N-n}}$ are the generalized $W$ or Dicke states \cite{wiegner2011quantum}. The latter are defined as
\begin{equation} \label{eq:Wstate}
    \ket{W_{n,N-n}}=\binom{N}{n}^{-1/2}\sum_k\mathcal{P}_k\ket{S_{n,N-n}}\;,
\end{equation}
where $\mathcal{P}_k$ is the operator that produces all the possible different permutations with equal number $n$ of particles being in the $\ket{g_+}$ state, and with $\ket{S_{n,N-n}}$ defined as
\begin{equation}
    \ket{S_{n,N-n}}=\prod_{\alpha=1}^ne^{i\phi_\alpha}\ket{g_+}_\alpha
    \prod_{\beta=n+1}^N\ket{g_-}_\beta\;.
\end{equation}
Heralded photon entanglement obtained via single photon detection represents the case with $n=1$. For example, the formula can be reduced to the two-ion case reported in Ref.~\cite{Slodička13} by setting the condition $N=2$ to obtain $\ket{\psi}=\left(\ket{g_-g_+}+e^{i\phi}\ket{g_+g_-}\right)/\sqrt{2}$, where the phase $\phi$ is determined by the difference in laser excitation phase and path lengths of the photons to the detection point, from different ions. The process is reliable in the limit of weak excitation probability ($p \ll 1$) because the infidelity is bounded by the events in which the detector clicks as a consequence of multi-photon emission if they cannot be experimentally separated from the single-photon events. Such separation might be possible by considering the spatial distribution of single and multiple photon events in Fourier space to further shape the sectors of the SLM to predominantly deflect the single photon contribution~\cite{Wolf20}. However, it is beyond the scope of this article to quantify the extent of such an advantage. In the scenario in which multiple and single photon events cannot be distinguished nor separated, a click in the detector heralds the creation of the $N$-particle entangled state $\ket{W_{1,N-1}}$, with a fidelity that depends on the probability of single photon emission $p$ via the function
\begin{equation}\label{eq:Cabrillo}
\displaystyle
    F=\frac{Np(1-p)^{N-1}}{\sum\limits_{n=1}^{N}\binom{N}{n}p^n(1-p)^{N-n}} \, ,
\end{equation}
which is a monotonically decreasing function of $p$. For computing purposes, when encoding a logical qubit on a 2D surface code, the fault-tolerance threshold for the fidelity is 99\%~\cite{Campbell2017}. This limit can be achieved with $p<0.01$ for $N=2$, and similar values for $p$ have already been achieved experimentally~\cite{Araneda18}.

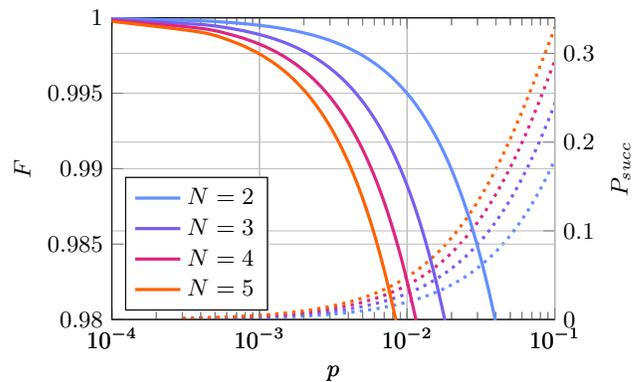
\begin{figure}[t!]
    \centering
    \begin{tikzpicture}
     \begin{semilogxaxis}[	
        width=\textwidth/2.4, height=\axisdefaultheight/1.3, 
        axis y line*=right,
        xmin = 0.0001, xmax = 0.1,
        ymin = 0.0, ymax = 0.34,
        xlabel = {$p$},
	ylabel = {$P_{succ}$},
        legend cell align = {left},
        grid=major,
	legend pos = south west,
        yticklabel style={scaled ticks=false,/pgf/number format/fixed,/pgf/number format/precision=3},]
    \addplot[smooth, very thick, color1, dotted] 
    file[skip first] {Figures/Entanglement/Psucc/p_succ2.dat};
    \addplot[smooth, very thick, color2, dotted] 
     file[skip first] {Figures/Entanglement/Psucc/p_succ3.dat};
    \addplot[smooth, very thick, color3, dotted] 
     file[skip first] {Figures/Entanglement/Psucc/p_succ4.dat};
    \addplot[smooth, very thick, color4, dotted] 
     file[skip first] {Figures/Entanglement/Psucc/p_succ5.dat};
\legend{$N=2$, 
	$N=3$,
	$N=4$,
        $N=5$}
\end{semilogxaxis}
     \begin{semilogxaxis}[	
        width=\textwidth/2.4, height=\axisdefaultheight/1.3, 
        axis y line*=left,
        xmin = 0.0001, xmax = 0.1,
        ymin = 0.98, ymax = 1,
        xlabel = {$p$},
	ylabel = {$F$},
        legend cell align = {left},
        grid=major,
	legend pos = south west,
        yticklabel style={scaled ticks=false,/pgf/number format/fixed,/pgf/number format/precision=3},]
    \addplot[smooth, very thick, color1,] 
    file[skip first] {Figures/Entanglement/Fidelity/fidelity_2.dat};
    \addplot[smooth, very thick, color2,] 
     file[skip first] {Figures/Entanglement/Fidelity/fidelity_3.dat};
    \addplot[smooth, very thick, color3,] 
     file[skip first] {Figures/Entanglement/Fidelity/fidelity_4.dat};
    \addplot[smooth, very thick, color4,] 
     file[skip first] {Figures/Entanglement/Fidelity/fidelity_5.dat};
\legend{$N=2$, 
	$N=3$,
	$N=4$,
        $N=5$}
\end{semilogxaxis}
    \end{tikzpicture}
    \caption{The maximum fidelity $F$ from Eq.~\eqref{eq:Cabrillo} (solid line) and the probability for single-photon events $P_{succ}$ defined in  Eq.~\eqref{eq:psucc} (dotted line) of the Cabrillo protocol versus the state preparation probability $p$, for different numbers of ions (solid line). We hereby assumed, that multiple and single photon events cannot be distinguished.}
    \label{fig:Cabrillo_fidelity}
\end{figure}

The probability of emitting and detecting only one photon $P_{succ}$ is the pre-factor of $\ket{W_{1,N-1}}$, given by the expression
\begin{equation}\label{eq:psucc}
    P_{succ}=\rho N p(1-p)^{N-1}. 
\end{equation}
Eq.~\eqref{eq:Cabrillo} and the success probability $P_{succ}$ versus $p$ for different numbers of ions are shown in Fig.~\ref{fig:Cabrillo_fidelity}. The losses for a detection apparatus featuring a 0.6 numerical aperture objective lens (10\% of light collection efficiency, 91.5\% optical transmission at \qty{493}{\nano\meter}) and an EMCCD detector (around $90\%$ of quantum efficiency at \qty{493}{\nano\meter}) amounts to $\epsilon_2=0.08$. The SLM we have considered in this article has a reflection coefficient of $0.98$ and its diffraction efficiency can be modeled as cross-talk between neighbouring pixels, resulting in an overall loss of $\epsilon_1=0.83$, i.e. $\rho\approx0.07$. Considering an experimental duty cycle of \qty{3}{\kilo\hertz} and $p=0.05$ we expect an entanglement generation rate of $\approx\qty{20}{\second^{-1}}$.

\subsection{\label{sec:collective-spontaneous-emission}Collective spontaneous emission}
The configuration depicted in Fig.~\ref{fig:optical_setup} with a flat mirror in place of the SLM is known as half-cavity~\cite{Dubin2007, Hetet2010, Dorner02} and can be used to control the SE of a single ion. With the SLM, the control can be extended to multiple ions located in the focal plane of the half-cavity. The SLM enables the control of the scattering rate of all ions simultaneously, so that their collective emission can be enhanced or suppressed. For instance, we can solve Eq.~\eqref{eq:field_detector_theta} imposing a suppression condition, i.e. the condition of destructive interference $\tilde{u}(\bm{k})=0$. The function $\tilde{f}\left(\bm{k}\right)$ can be separated into amplitude $A(\bm{k})$ and phase $\phi(\bm{k})$ according to the formula: $\tilde{f}(\bm{k}) = A\left(\bm{k}\right)e^{i\phi\left(\bm{k}\right)}\;$. If in the proposed setup, the function $f\left(\bm{r}\right)$ is real, the following relation holds $\tilde{f}\left(-\bm{k}\right) =A\left(\bm{k}\right)e^{-i\phi\left(\bm{k}\right)}$. This condition is equivalent to stating that all emitters are in phase, which can be achieved experimentally by driving the ions with a laser beam orthogonal to the chain. In a few steps we arrive to the following condition for the SLM's phase mask
\begin{equation}
    \label{eq:invisibility_mask}
    \tilde{m}\left(\bm{k}\right) = \exp\left(-i2\phi\left(\bm{k}\right)\right) \; .
\end{equation}
By inserting this expression into Eq.~\eqref{eq:field_chain_summed}, we obtain the field in the imaging plane
\begin{equation}
    \label{eq:invisibility_space}
    u(\bm{r})=f(\bm{r})\left(1-\rho e^{i\psi}\right) \; .
\end{equation}
The field $f(\bm{r})$ is unaltered in shape, but is modulated by the term $(1-\rho e^{i\psi})$. This implies that, even if the emitters remain distinguishable in space in the imaging plane, the phase mask influences the total field at the detector. Enhancement and suppression of the amplitude can be obtained by changing the distance between the SLM and $L_1$ to control the global phase $\psi$.

The result derived in Eq.~\eqref{eq:invisibility_space} considers only the interference of the fields, assuming that the ions scatter phase-coherent monochromatic radiation. To assess the validity of this approximation, we have refined our model by incorporating the main factors that limit the light coherence of such quantum scatterers, and thus the contrast of interference. The first effect is that ions can become saturated under a strong driving field, causing them to scatter light without spatial coherence~\cite{Wolf20}. The second effect to consider is that each ion acts as a photon source, and its temporal coherence is limited by the lifetime of the electronic state. The third source of decoherence arises from the harmonic motion of the ions within the trapping potential. To compare the three contributions to the residual light intensity we calculate the dimensionless quantities $C_1$, $C_2$, and $C_3$. The coefficients are the intensity of each ion's residual image normalized by the individual ion free-space image as if the SLM was not present. The expression of each $C$ coefficient can be understood as if only one of the three effects is active at a time. The three contributions can be then additively combined as presented in the Appendices.

We simulate these effects with a numerical model that aims at evaluating the minimum attainable visibility, i.e. under the assumptions $\rho=1$ and $\psi=0$, where the wavefront gets reflected back with opposite phase and destructive interference with the original field happens. We assume the system to be composed of $N$ ions localized at locations $\bm{r}_n$. For simplicity, we follow the approximation already applied in the literature that assumes the ions to be two-level systems with an excited state decay rate $\gamma$~\cite{Skornia01}. The ions are driven with uniform Rabi frequency $\Omega$ and a detuning $\Delta$ and we explain the model in its entirety in Appendix~\ref{sec:app_driven}.

In our model, the first contribution stemming from saturation can be identified in the free space far-field intensity distribution $I(\bm{k})$ measurable by a detector, that can be written as the sum of a ``coherent'' and an ``incoherent'' intensity: $I(\bm{k}) = I_c(\bm{k}) + I_i(\bm{k})$. The two intensities have the expressions 
\begin{align}
    \label{eq:coherent_intensity}
    I_c\left(\bm{k}\right)= & \left(\frac{\Omega \sqrt{\Delta^2+\gamma^2}}{\Omega^2+2\Delta^2+2\gamma^2}\right)^2 \abs{ \sum_n e^{i \bm{k} \cdot \bm{r}_n } }^2\\
    \label{eq:incoherent_intensity}
    I_i\left(\bm{k}\right)= &\frac{N}{2} \left(\frac{\Omega^2}{\Omega^2+2\Delta^2+2\gamma^2}\right)^2,
\end{align}
and they are derived in detail in Eq.~\eqref{eq:farfield_noSLM}. This result is compatible with what was found in former literature for the emission of light from two ions ($N=2$)~\cite{Skornia01}. We note that this formula relates to the field function $f$ previously introduced via the relation $I_c\left(\bm{k}\right) \propto \lvert\tilde{f}\left(\bm{k}\right)\rvert^2$, which shows that the mask reported in Eq.~\eqref{eq:invisibility_mask} is calculated considering only the coherent part of the emitted field and it will not be able to cancel the incoherent contribution. The ions remain distinguishable in the imaging plane and their light intensity is homogeneously modulated by changing the phase $\psi$. However, the minimum attainable light intensity, even with unit reflectivity, is not zero. The remaining images of the ions are stemming from the incoherent term $I_i\left(\bm{k}\right)$. This implies that such images, being fully incoherent, can no longer interfere with each other. We define the first coefficient $C_1$ that captures the contribution of saturation to the residual image as follows
\begin{equation}\label{eq:c_1}
    C_1(\gamma, \Omega, \Delta) := \frac{I_i\left(\bm{k}\right)}{N}= \frac{\Omega^4}{2(\Omega^2 + 2\Delta^2 + 2\gamma^2)^2} \; .
\end{equation}

\begin{figure}[t!]
    \centering
    \begin{tikzpicture}
	\begin{axis}[	
 width=\textwidth/2.1, height=\axisdefaultheight/1.3, 
 xmin=0, xmax=10, ymin=0,
            xlabel = {$\Omega/\gamma$},
	    ylabel = {$C_1(\tau)$},
     	legend cell align = {left},
            grid=major,
	      legend pos = north east,]
     
    \addplot[
	    smooth,
	    very thick,
	    color1,] 
     file[skip first] {Figures/Invisibility/Saturation/scor1.dat};

     \addplot[
	    smooth,
	    very thick,
	    color2,] 
     file[skip first] {Figures/Invisibility/Saturation/scor2.dat};

     \addplot[
	    smooth,
	    very thick,
	    color3,] 
     file[skip first] {Figures/Invisibility/Saturation/scor3.dat};

\legend{
	$\Delta=0$, 
	$\Delta=\gamma$,
	$\Delta=5\gamma$
}
\end{axis}
    \end{tikzpicture}
    \caption{The function $C_1$ as defined in Eq.~\eqref{eq:c_1} modulating the mean intensity $I^{\mathrm{det}}$ remaining on the detector, for different detunings $\Delta$.}
    \label{fig:c1}
\end{figure}
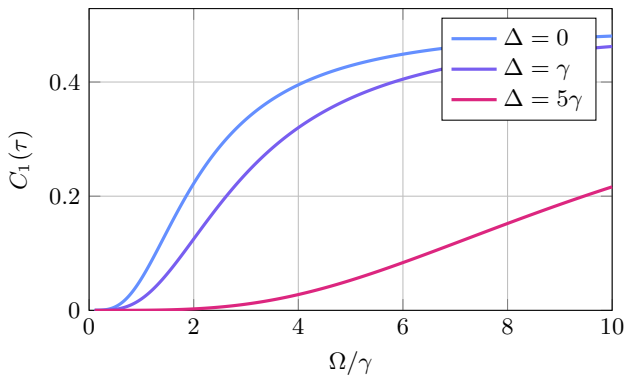

This coefficient determines the residual intensity as a function of the model parameters and it is not dependent on the number of ions. We note that in the case of weak driving it is suppressed to fourth order in $\Omega$. The value of $C_1$ is plotted in Fig.~\ref{fig:c1} as a function of $\Omega/\gamma$ for different detunings $\Delta$. In practical experiments that observe light scattering of trapped ions, the driving laser has a detuning $\Delta$ of maximum a few times $\gamma$ and a driving strength $\Omega/\gamma$ around one~\cite{cerchiari2021one}. From the plot, we evince that only small residual intensity ($C_1\lesssim0.1$) are present. Furthermore, existing experiments that would implement the SLM setup could already achieve a measurement of the variation of $C_1$ while transitioning between the unsaturated and the saturated regime.

Secondly, the contribution $C_2$ due to temporal incoherence stemming from the lifetime is connected to the time $\tau$ spent by the light to go and return from the SLM. The interference contrast reduces if $\tau$ is comparable to or longer than the excited state lifetime~\cite{Dorner02}. The proportionality coefficient $C_2$ for the residual intensity at the detector is given by the expression
\begin{equation}\label{eq:c_2}
    C_2(\tau,\gamma,\Omega,\Delta) = \ev{\sigma_{+,H}(\tau) \sigma_-} - \ev{\sigma_+ \sigma_-}\,.
\end{equation}
In this expression, $\sigma_\pm$ are level raising and lowering operators for a single ion and the $\tau$-dependent correlator is evaluated in Eq.~\eqref{eq_app:tau_correlator}.
The quantity $C_2$ is given in units of the free-space intensity of a single ion arriving at the detector. In realistic experimental conditions~\cite{cerchiari2021one} and by placing the SLM at $\qty{0.1}{\meter}$ from the ions, one has $C_2\approx\qty{2.5e-6}{}$ ($\tau\gamma=0.027$). This contribution to decoherence is therefore negligible in most practical applications. An extended description of coefficient $C_2$, with an alternative derivation of $C_1$, are reported in Appendix~\ref{app:sat_tempcoh}.

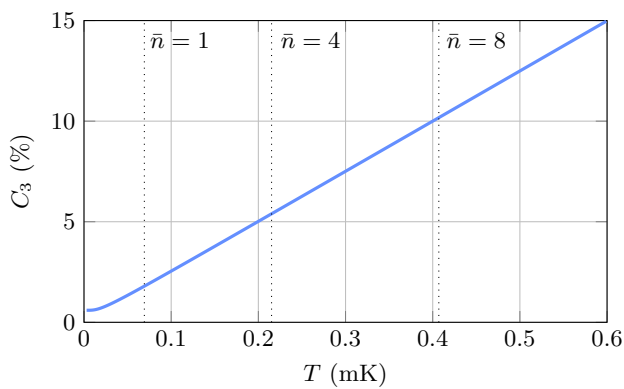
\begin{figure}[t!]
    \centering

    \begin{tikzpicture}
	\begin{axis}[	
 width=\textwidth/2.1, height=\axisdefaultheight/1.3, 
 xmin=0, xmax=0.6,ymin=0,ymax=15,
            xlabel = $T$ (mK),
	    ylabel = {$C_3$} (\%),
     	legend cell align = {left},
            grid=major,
	      legend pos = north east,]
     
    \addplot[smooth, very thick, color1,] file[skip first] {Figures/Invisibility/Motion/C_motion.dat};
     \draw[color=black, dotted] (axis cs:0.0692335, 0) -- (axis cs:0.0692335, 15);
     \node[] at (axis cs: 0.11,14) {$\bar{n}=1$};
     \draw[color=black, dotted] (axis cs:0.21515, 0) -- (axis cs:0.21515, 15);
     \node[] at (axis cs: 0.26,14) {$\bar{n}=4$};
     \draw[color=black, dotted] (axis cs:0.407, 0) -- (axis cs:0.407, 15);
     \node[] at (axis cs: 0.45,14) {$\bar{n}=8$};
\end{axis}
    \end{tikzpicture}
    
    \caption{Percentage factor $C_3$, representing the remnant intensity due to the motion of the ion vs the ion temperature $T$. In this plot, $T=0$ K correspond to $\bar{n}=0$ and $T=0.6$ mK to $\bar{n}\approx12$ (Doppler limit). Compared to the other two contributions, this represents the largest one to decoherence and loss of contrast in the interference.}
    \label{fig:c3}
\end{figure}

We consider now the third and largest contribution to decoherence, which results from the ion motion. Modern SLMs permit to refresh the phase function in the millisecond timescale. This update speed is much slower than the ion oscillation frequency inside the trap which is typically about $\omega = 2 \pi \times $~\SI{1}{\mega\hertz}. Therefore, the phase mask must be selected based on the average position of the ions. Consequently, fluctuations of the ions' position results in improper match between the direct and reflected fields that are interfering at the detector. To assess how the mismatch affects the final contrast, we consider that the positions of the ions are randomly distributed according to a Gaussian thermal oscillator state, expressing the field as $f(\bm{r}_n) = f_0(\bm{r}_n) + \delta f(\bm{r}_n)$.
In the limit of small motional amplitudes (compared to the wavelength), the random fluctuations have zero mean, $\langle \delta f(\bm{r}_n) \rangle = 0$, and can be assumed uncorrelated between different ions $\langle \delta f(\bm{r}_n) \delta f(\bm{r}_m) \rangle = 0$ for all $m \neq n$; see derivation in Appendix~\ref{sec:app_ionmotion}.

In an isotropic ion trap potential of frequency $\omega$, the thermal standard deviation of each ion's position is $\sigma = \sqrt{(2\bar{n}+1)\hbar/2m \omega}$, where $m$ is the ion mass, and $\bar{n} = (e^{\hbar \omega / k_B T}-1)^{-1}$ is the mean phonon number.
Assuming a Gaussian-shaped single-ion image with a diffraction-limited spot size $s$, we can expand $C_3$ to lowest order in $\sigma$ as
\begin{equation}
    C_3 =\left(\frac{\sigma}{2s}\right)^2+\left(k\sigma\right)^2\;,
\end{equation}
where $k=2\pi/\lambda$. Here, we take a Gaussian single-ion image of characteristic width $s$ and isotropic position fluctuations with scale $\sigma$. The dependence of $C_3$ as a function of temperature is presented in Fig.~\ref{fig:c3} for the mass value of Barium: $m=138$~amu. We see that at the Doppler limit $\bar{n}\approx8-10$ the contrast is reduced in a similar way as observed experimentally in experiments with a single ion and a mirror~\cite{Slodička2012}. To obtain the maximum contrast, further cooling techniques are required.
\section{\label{conclusions} Conclusions}
In this article, we describe how to control the SE of single and multiple light scatterers such as trapped ions with a phase-programmable mirror as boundary condition. We presented two applications: how this setup can be used for programmable photon-mediated entanglement and for controlling the SE of distinguishable ions.

Programmable entanglement operations mediated by light can be adopted for on-chip preparation of the entangled state of trapped ions in near-future quantum processors. Light-based operations have the potential to reduce the shuttling budget on chip for the preparation of the initial entangled state of trapped ions which are located at any trapping site. The proposed setup can be built without moving parts and is complementary to other approaches based on integrated photonics~\cite{Mehta2020, Jiang2011}. Compared to waveguide-based interconnections, however, our design adapts to a large variety of trap designs and imposes no constraints on the location of the ions on the trap structure.

The SLM enables modulation of the ions' spontaneous emission while maintaining the spatial identity of each individual ion. This is a step forward towards understanding how to control the visibility of complex objects because, compared to state-of-the-art approaches, it does not require emitters to be made indistinguishable~\cite{Obsil19}, meaning that in the imaging plane the ions appear as separated light sources. The proposed method modifies the emission at a distance, leaving room for the experimenter to manipulate the sample on the opposite side of the SLM and, thus, operates without imposing stringent conditions on the sample under investigation~\cite{Busch1998}. The suppression of spontaneous emission for multiple ions can also find applications in reducing the off-resonant scattering error of certain quantum gates, such as Light Shift gates~\cite{Ozeri07}, that rely on the weak excitation of a dipole allowed transition. Furthermore, the technique does not strictly require the sample to be made of trapped ions, but only that the sample can scatter light. For this reason, we believe that it can find application in the field of microscopy for affecting the visibility of nearby scatterers globally or differentially.

\begin{acknowledgments}
This work was supported by the Ministry of Culture and Research of the State of North Rhine-Westphalia, by the Austrian Science Fund (FWF) project number: P 36233-N (SONATINA) and by the Institut für Quanteninformation GmbH. G.A. acknowledges support from Wolfson College Oxford. The authors would like to thank Dr.~Alex Erhard, Dr. Lukáš Slodička and Dr. Maurizio Verde for fruitful discussions.
\end{acknowledgments}

\appendix

\section{\label{sec:appendix-detailed-model}{Model description}}
We report in this section a more detailed derivation of the formulas presented in Sec.~\ref{sec:setup}. Considering the arrangement depicted in Fig.~\ref{fig:optical_setup}, the ions are located in a single plane ($x,y,z=0$) corresponding to the focal plane of the lenses $L_1$ and $L_2$. The function $f(\bm{r})$, where $\bm{r}=(x,y,z)$, describes the electric field of the emitters. The electric field at another point $\bm{r}'$ in space will be given by convolution of $f$ with the far-field Green's function of the Huygens-Fresnel integral~\cite{NovotnyHecht2012}
\begin{equation}
\label{eq_app:convolution_free_space}
    f(\bm{r}')=\frac{1}{2i\lambda}\int d^2\bm{r}\,\, f(\bm{r})\frac{e^{ik|\bm{r}-\bm{r}'|}}{|\bm{r}-\bm{r}'|}.
\end{equation}
In far-field approximation $|\bm{r}'|\gg|\bm{r}|$, the electric field takes the form
\begin{equation}
    f(\bm{r}')=\frac{1}{2i\lambda}\frac{e^{ik| \bm{r}'|}}{|\bm{r}'|}\int d^2\bm{r}\,\, f(\bm{r})e^{-ik|\bm{r}|\hat{\bm{r}}\cdot\hat{\bm{r}}'},
\end{equation}
in which we simplified $|\bm{r}-\bm{r}'|\approx|\bm{r}'|$ in the denominator. In the prefactor, we can neglect the imaginary terms corresponding to a shift in the phase and the constants factors including $\lvert\bm{r}'\lvert$. It is also convenient to rewrite the exponent as $k|\bm{r}|\hat{\bm{r}}\cdot\hat{\bm{r}}'=\bm{k}\cdot\bm{r}$ and note that the resulting integral is the Fourier transform of $f(\bm{r})$
\begin{equation}
    \label{eq_app:first_transform}
    \tilde{f}(\bm{k})=\int d^2\bm{r}\,\,f(\bm{r})e^{-i\bm{k} \cdot \bm{r}}.
\end{equation}
Via a flat mirror and a lens, we reflect the field back into the $(xy)$ plane. This corresponds to performing an additional FT
\begin{equation}
    \label{eq_app:second_transform}
    f(-\bm{r})=\int d^2\bm{k}\,\,\tilde{f}(\bm{k})e^{-i\bm{k}\cdot\bm{r}}.
\end{equation}
We now consider the radiation to be reflected with a SLM instead of a mirror. The Green's function is modified by the multiplicative phase function $\tilde{m}(\bm{k})$ generated on the SLM's pixels. By virtue of the convolution theorem for Fourier transforms, the reflected field amplitude becomes
\begin{align}
\label{eq_app:SLM_reflection}
    f_m(-\bm{r})&=\int d^2\bm{k}\,\,\tilde{m}(\bm{k})\tilde{f}(\bm{k})e^{-i\bm{k}\cdot\bm{r}}\nonumber=(f\ast m)(-\bm{r}).
\end{align}
Consistently, the result reduces to Eq.~\eqref{eq_app:second_transform} if we assume reflection at a perfect mirror by setting $\tilde{m}=1$. This means that if the emitter is displaced by $\bm{r}$ from the optical axis, its image will be formed in its same plane but at position $-\bm{r}$. The actual field on the emitters' plane is found by summing the one emitted directly and the one reflected by the SLM and has expression
\begin{equation}
     \label{eq_app:total_field_in_emitter_plane}
     u(\bm{r})=f(\bm{r})+f_m(\bm{r})=f(\bm{r})+\rho e^{i\psi}(f\ast m)(-\bm{r}).
\end{equation}

In the last equation, the non-ideal reflectivity $\epsilon_1$ of the device and the optical losses $\epsilon_2$ through the imaging system are incorporated in the $\rho=\epsilon_1\epsilon_2$ coefficient. Also considered, is an additional constant phase term emerging from the optical path of the light traveling to and from the light modulator $\psi=\omega\Delta t$, where $\omega$ is the frequency of the scattered light and $\Delta t$ the round-trip time. The field $u(\bm{r})$ is propagated in free space and under paraxial approximation takes the form
\begin{align}
    \label{eq_app:sum_field_propagated}
    \tilde{u}(\bm{k}')&=\int d^2\bm{r}\,\,u(\bm{r})e^{-i\bm{k}'\cdot\bm{r}}=\int d^2\bm{r}\,\,\bigl[f(\bm{r})+\rho e^{i\psi}\nonumber\\\cdot\int &d^2\bm{r}_p\,\,m(-\bm{r}-\bm{r}_p)f(\bm{r}_p)\bigr]e^{-i\bm{k}'\cdot\bm{r}}\nonumber\\
    &=\tilde{f}(\bm{k}')+\rho e^{i\psi}\tilde{m}(-\bm{k}')\int d^2\bm{r}_p\,\,f(\bm{r}_p)e^{+i\bm{k}'\cdot\bm{r}_p}\nonumber\\
    &=\tilde{f}(\bm{k}')+\rho e^{i\psi}\tilde{m}(-\bm{k}')\tilde{f}(-\bm{k}')\; .
\end{align}
Finally, the impinging field on the detector $d(\bm{r}')$ is found by inverting the argument of Eq.~\eqref{eq_app:total_field_in_emitter_plane} and performing a change of variable $\bm{r}\to\bm{r}'$ for the sake of clarity.

\section{Coherently driven two-level ions} \label{sec:app_driven}
Here, we consider a model of $N$ coherently driven two-level ions in order to calculate the far-field emitted intensity and to show that the method for suppressing spontaneous emission can equally well be used for stimulated emission.
This can be considered a generalization of the two-ion model in Ref.~\cite{Skornia01}.

The ions are driven by a classical field of frequency $\omega_L$ and polarization $\bm{e}_L$.
They are assumed to be widely spaced with respect to the wavelength $\lambda_0 = 2\pi c/\omega_0$ of their transition, such that they are non-interacting and their spontaneous emission is independent.
Following a textbook approach~\cite{GardinerZoller_ch14}, the dynamics of a set of driven two-level ions in free space coupled to the electromagnetic field are governed by the Markovian master equation
\begin{align} \label{eq_app:master}
     \frac{\dd \rho}{\dd t} = -\frac{i}{\hbar} \left[ H_a + H_d(t), \rho \right] + \sum_{n=1}^N 2\gamma \mathcal{D}[\sigma_-^{(n)}] \rho,
\end{align}
where $\rho(t)$ is the quantum state of the $N$ atoms at time $t$ in the Schr\"odinger picture, $H_a = -\frac{1}{2} \hbar \omega_0 \sum_{n=1}^N  \sigma_z^{(n)}$ is the free Hamiltonian of the ions, $H_d(t) = \frac{\hbar \Omega}{2} \sum_{n=1}^N [e^{-i\omega_L t} \sigma_+^{(n)} + e^{i\omega_L t} \sigma_-^{(n)}]$ is the driving Hamiltonian with Rabi frequency $\Omega$, and $\mathcal{D}[\sigma_-^{(n)}]\rho = \sigma_-^{(n)} \rho \sigma_+^{(n)} - \frac{1}{2}(\sigma_+^{(n)}\sigma_-^{(n)} \rho + \rho \sigma_+^{(n)} \sigma_-^{(n)})$ is the dissipator describing spontaneous emission of ion $n$ with rate coefficient $\gamma$.

We start by finding the steady state of the ions, reached in the long-time limit.
Note that the master equation is separable, so each ion's state $\rho^{(n)}$ obeys the same independent dynamics.
In the interaction picture rotating at frequency $\omega_L$, $\rho^{(n)}_I := e^{-i\omega_L t\sigma_z/2} \rho^{(n)}e^{i\omega_L t \sigma_z/2}$ evolves as
\begin{align} \label{eq_app:master_interaction}
    \frac{\dd \rho_I^{(n)}}{\dd t} = -\frac{i \Delta}{2} [\sigma_z, \rho_I^{(n)}] -\frac{i\Omega}{2}[\sigma_x, \rho_I^{(n)}] + 2\gamma \mathcal{D}[\sigma_-]\rho_I^{(n)},
\end{align}
where $\Delta = \omega_L - \omega_0$ is the detuning.
The time-independent steady state in this frame can be written as $\rho_\infty^{(n)} = \frac{1}{2}(I + r_+ \sigma_+ + r_- \sigma_- + r_z \sigma_z)$, with
\begin{align}
    r_+ = r_-^* & = \frac{2\Omega(\Delta - i\gamma)}{\Omega^2 + 2\Delta^2 + 2\gamma^2}, \\
    r_z & = \frac{2(\Delta^2 + \gamma^2)}{\Omega^2 + 2\Delta^2 + 2\gamma^2}.
\end{align}
The full steady state of all ions in the interaction picture is then $\rho_\infty= \bigotimes_n \rho_\infty^{(n)}$.

The emitted component of the field $\bm{E}^\text{out}(\bm{r},t)$ in the Heisenberg picture can be found from the input-output formalism to have positive-frequency component
\begin{align} \label{eq_app:field_out}
    \bm{E}_+^\text{out}(\bm{r},t) = \kappa \sum_n \frac{\bm{e}_{\perp n}}{s_n} \sigma_{-,H}^{(n)}\left( t - \frac{s_n}{c}\right),
\end{align}
where $\bm{s}_n = \bm{r} - \bm{r}_n$ is the vector of propagation from the position of the $n$th ion to $\bm{r}$, and $\bm{e}_{\perp n}$ is the component of $\bm{e}_L$ perpendicular to $\bm{s}_n$. Here, $\kappa$ is a constant related to $\gamma$ whose value we will not need. Note that the symbol $s_n$ indicates
\begin{equation}
    \label{eq:sn_roman}
    s_n = \sqrt{\lvert \bm{r}\rvert^2 + \lvert \bm{r}_n\rvert^2 - 2 \bm{r} \cdot \bm{r}_n} \approx \lvert \bm{r}\rvert - \hat{\bm{r}} \cdot \bm{r}_n
\end{equation}
where the last step is valid in far field approximation.
The steady-state intensity $\ev*{\bm{E}_-^\text{out}\cdot\bm{E}_+^\text{out}}$ at position $\bm{r}$ can be found by noting that $\ev*{\sigma_{+,H}^{(m)}(t)\sigma_{-,H}^{(n)}(0)} = \ev*{\sigma_{+,H}^{(m)}(t)}\ev*{\sigma_{-,H}^{(n)}(0)}$ for $m \neq n$ due to the lack of correlations in the steady state.
Then, using the interaction picture, 
\begin{align}
    \ev*{\sigma_{-,H}^{(n)}(t)} & = \tr[e^{-i\omega_L t \sigma_z/2} \sigma_-^{(n)} e^{i\omega_L t \sigma_z/2} \rho_\infty^{(n)}] \nonumber \\
    & = e^{-i\omega_L t} \tr[\sigma_-^{(n)} \rho_\infty^{(n)}] \nonumber \\
    & = \frac{e^{-i\omega_L t} r_+}{2},
\end{align}
so we have
\begin{align} \label{eq_app:steady_expvals}
    \ev*{\sigma_{+,H}^{(m)}(t - s_m/c) \sigma_{-,H}^{(n)}(t-s_n/c)} = \\
    \quad
        \begin{cases}
        \frac{\Omega^2}{2(\Omega^2 + 2\Delta^2 + 2\gamma^2)} & m=n, \nonumber \\
        \frac{\Omega^2(\Delta^2+\gamma^2)}{(\Omega^2 + 2\Delta^2 + 2\gamma^2)^2} e^{i\omega_L(s_n-s_m)/c} & m \neq n
    \end{cases}.
\end{align}

\begin{figure}[h!]
    \centering
    \begin{tikzpicture}
	\begin{axis}[
 width=\textwidth/2.2, height=\axisdefaultheight/1.3, 
            xmin = 0.0, xmax = 10,
            ymin = 0.0, ymax = 0.5,
            xlabel = {$\Omega/\gamma$},
	    ylabel = {$\lvert\bra{1}\rho_\infty\ket{1}\lvert$},
     	legend cell align = {left},
            grid=major,
	      legend pos = south east,]
     
    \addplot[
	    smooth,
	    very thick,
	    color1,] 
     file[skip first] {Figures/Invisibility/Population/population_1.dat};

     \addplot[
	    smooth,
	    very thick,
	    color2,] 
     file[skip first] {Figures/Invisibility/Population/population_2.dat};

    \addplot[
	    smooth,
	    very thick,
	    color3,] 
     file[skip first] {Figures/Invisibility/Population/population_3.dat};

\legend{
	$\Delta=0$, 
	$\Delta=\gamma$,
	$\Delta=5\gamma$
}
\end{axis}
    \end{tikzpicture}
    \begin{tikzpicture}
	\begin{axis}[	
 width=\textwidth/2.2, height=\axisdefaultheight/1.3, 
            xmin = 0.0, xmax = 10,
            ymin = 0.0, ymax = 0.38,
            xlabel = {$\Omega/\gamma$},
	    ylabel = {$\lvert\bra{0}\rho_\infty\ket{1}\lvert$},
     	legend cell align = {left},
            grid=major,
	      legend pos = south east,]
     
    \addplot[
	    smooth,
	    very thick,
	    color1,] 
     file[skip first] {Figures/Invisibility/Coherence/coherence_1.dat};

     \addplot[
	    smooth,
	    very thick,
	    color2,] 
     file[skip first] {Figures/Invisibility/Coherence/coherence_2.dat};

    \addplot[
	    smooth,
	    very thick,
	    color3,] 
     file[skip first] {Figures/Invisibility/Coherence/coherence_3.dat};

\legend{
	$\Delta=0$, 
	$\Delta=\gamma$,
	$\Delta=5\gamma$
}
\end{axis}
    \end{tikzpicture}
    \caption{Properties of the single-atom steady state $\rho_\infty$ in the continuously driven model as a function of driving strength $\Omega$ for different detunings $\Delta$: a) excited state population (top) and coherence (bottom).}
    \label{fig:driven_model}
\end{figure}
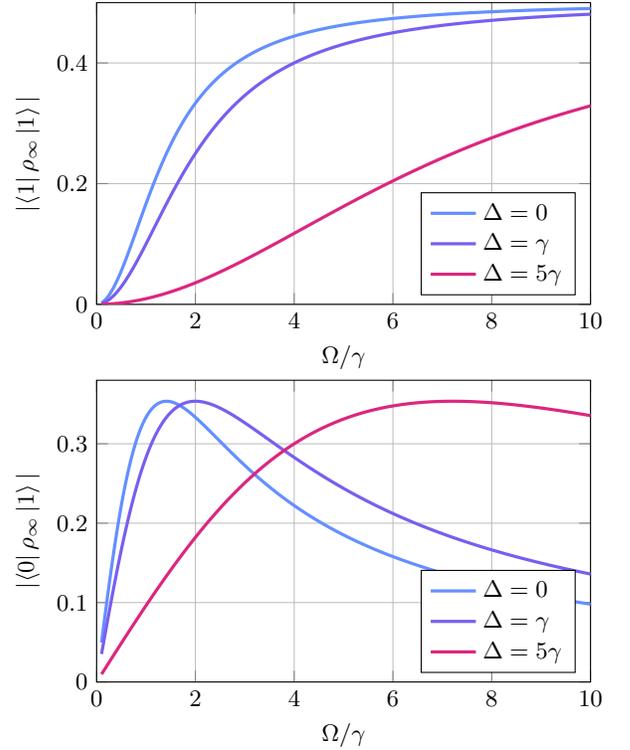

The excited state population and coherence of the steady state are plotted in Fig.~\ref{fig:driven_model}.

We now assume a paraxial approximation and a small total spatial extent of the set of ions, which allows us to simplify $\bm{e}_{\perp n} \approx \bm{e}_\perp$ independently of the ion, and similarly the factor of $s_n$ in the denominator of Eq.~\eqref{eq_app:field_out} is replaced by a constant distance $R$ from the ions to the detector.
Then Eq.~\eqref{eq_app:field_out} reduces to
\begin{align} \label{eq_app:field_out_parax}
    \bm{E}_+^\text{out} \approx \frac{\kappa}{R} \bm{e}_\perp \sum_n \sigma_{-,H}^{(n)}\left(t - \frac{s_n}{c} \right),
\end{align}
and from Eq.~\eqref{eq_app:steady_expvals}, the mean intensity is
\begin{align}
    \ev{\bm{E}_-^\text{out} \cdot \bm{E}_+^\text{out}} & \approx \frac{\kappa^2}{R^2} \abs{\bm{e}_\perp}^2 \left[ \frac{\Omega^2}{2(\Omega^2+2\Delta^2+2\gamma^2)} \cdot N + \right. \nonumber \\
        & \hspace{-5em} \left. \frac{\Omega^2(\Delta^2+\gamma^2)}{(\Omega^2+2\Delta^2+2\gamma^2)^2} \cdot \sum_{m<n} 2 \cos(\omega_L[s_n-s_m]/c) \right].
\end{align}
We can separate this into coherent and incoherent parts by writing
\begin{align}
    \label{eq:farfield_noSLM}
    \ev{\bm{E}_-^\text{out} \cdot \bm{E}_+^\text{out}} & \approx \frac{\kappa^2}{R^2} \abs{\bm{e}_\perp}^2 \left[ \abs{\frac{\Omega \sqrt{\Delta^2+\gamma^2}}{\Omega^2+2\Delta^2+2\gamma^2} \sum_n e^{i\omega_L s_n/c}}^2 \right. \nonumber \\
    & \left. + \frac{N}{2} \left(\frac{\Omega^2}{\Omega^2+2\Delta^2+2\gamma^2}\right)^2 \right] 
\end{align}
We therefore see that the coherent interference pattern dominates at weak driving, when $\Omega^2 \ll \Delta^2 + \gamma^2$. The equations \eqref{eq:coherent_intensity} and \eqref{eq:incoherent_intensity} of the main text are obtained combining Eqs.~\eqref{eq:farfield_noSLM} and~\eqref{eq:sn_roman}.\\

\section{Saturation and temporal coherence\label{app:sat_tempcoh}}
We now show that suppression of emission in the driven model is achievable using the SLM, with some residual field of two different origins:
one is an incoherent term due to saturation at strong driving,
and the other is due to finite coherence over the time delay $\tau$ in propagation from the ions to the detector via the SLM versus directly to the detector.

We denote the field components from emission in these two opposite directions along the $z$-axis by $\bm{E}^\text{SLM}$ and $\bm{E}^\text{dir}$, respectively.
Due to the placement of the ions in the $x$-$y$ plane, we can assume by symmetry that these take the same form, determined by Eq.~\eqref{eq_app:field_out_parax}.
The total field in the detector plane is then a sum of the direct and SLM reflected components
\begin{align}
    \bm{E}^\text{det}(\bm{r},t) = \bm{E}^\text{dir}(\bm{r},t) + \bm{E}^\text{SLM}(\bm{r},t-\tau),
\end{align}
with
\begin{align}
    \vb*{E}^\mathrm{dir}_+(\vb*{r},t) & = \sum_n \vb*{f}_n(\vb*{r}) \sigma_{-,H}^{(n)}(t - s_n/c) \nonumber, \\
    \vb*{E}^\mathrm{SLM}_+(\vb*{r},t-\tau) & = \rho e^{i\psi} \sum_n \vb*{g}_n(\vb*{r}) \sigma_{-,H}^{(n)}(t - s_n/c - \tau),
\end{align}
where $s_n$ is the total optical path length from ion $n$ to the detector, and $\bm{f}_n(\bm{r})$ encodes the spatial field variation in the image plane from the $n$th ion.
Here, $\vb*{g}_n(\vb*{r}) = (m * \vb*{f}_n)(-\vb*{r})$ is the reflected image of a single ion.
The SLM phase is chosen such that the total reflected field matches the direct field, i.e., such that $\vb*{G} := \sum_n \vb*{g}_n e^{i\omega_L s_n/c} = \sum_n \vb*{f}_n e^{i\omega_L s_n/c} =: \vb*{F}$.
The mean detected intensity is then $I^\mathrm{det} = \ev{\vb*{E}_+^\mathrm{det} \cdot \vb*{E}_-^\mathrm{det}} =  I^\mathrm{dir} + I^\mathrm{SLM} + 2\Re I^\mathrm{int}$ (dropping the explicit dependence on $\vb*{r}$ and $t$), where
\begin{align}
    I^\mathrm{dir} & = \sum_{m,n} \vb*{f}_m^\dagger \vb*{f}_n \ev{\sigma_{+,H}^{(m)}(t-s_m/c) \sigma_{-,H}^{(n)}(t-s_n/c)} \nonumber \\
        & = \sum_n \abs{\vb*{f}_n}^2 \ev{\sigma_+ \sigma_-}  \nonumber \\
        & \quad + \sum_{m \neq n} \vb*{f}_m^\dagger \vb*{f}_n e^{i \omega_L(s_n-s_m)/c} \abs{\ev{\sigma_-}}^2, \nonumber \\
        & = \abs{\sum_n \vb*{f}_n e^{i\omega_L s_n/c} \ev{\sigma_-}}^2 \nonumber \\
        & \quad + \sum_n \abs{\vb*{f}_n}^2 \left( \ev{\sigma_+ \sigma_-} - \ev{\sigma_+} \ev{\sigma_-} \right) \nonumber \\
        & = \abs{\vb*{F}}^2 \abs{\ev{\sigma_-}}^2 + \sum_n \abs{\vb*{f}_n}^2 \left( \ev{\sigma_+ \sigma_-} - \abs{\ev{\sigma_-}}^2 \right), \nonumber \\
        I^\mathrm{SLM} & = \rho^2 \abs{\vb*{G}}^2 \abs{\ev{\sigma_-}}^2 + \rho^2 \sum_n \abs{\vb*{g}_n}^2 \left( \ev{\sigma_+ \sigma_-} - \abs{\ev{\sigma_-}}^2 \right),
\end{align}
and the interference term is
\begin{align}
    I^\mathrm{int} & = \rho e^{i\psi} \sum_{m,n} \vb*{f}_m^\dagger \vb*{g}_n \ev{\sigma_{+,H}^{(m)}(t-s_m/c) \sigma_{+,H}^{(n)}(t-s_n/c-\tau)} \nonumber \\
        & = \rho e^{i\psi} \sum_n \vb*{f}_n^\dagger \vb*{g}_n \ev{\sigma_{+,H}(\tau) \sigma_-} \nonumber \\
        & \quad + \rho e^{i\psi} \sum_{m \neq n} \vb*{f}_m^\dagger \vb*{g}_n e^{i \omega_L(s_n/c - s_m/c + \tau)} \abs{\ev{\sigma_-}}^2 \nonumber \\
        & = \rho e^{i(\psi+ \omega_L \tau)} \vb*{F}^\dagger \vb*{G} \abs{\ev{\sigma_-}}^2 \nonumber \\
        & \quad + \rho e^{i(\psi + \omega_L \tau)} \sum_n \vb*{f}_n^\dagger \vb*{g}_n \left( \ev{\sigma_{+,H}(\tau) \sigma_-} - \abs{\ev{\sigma_-}}^2 \right) .
\end{align}
For these expressions, we used the lack of correlations in the steady state and Eq.~\eqref{eq_app:steady_expvals}.
Grouping these parts together, we have
\begin{align} \label{eqn:i_det1}
    I^\mathrm{det} & = \abs{\vb*{F} + \rho e^{i(\psi + \omega_L \tau)} \vb*{G}}^2 \abs{\ev{\sigma_-}}^2 \nonumber \\
        & \quad + \sum_n \abs{ \vb*{f}_n + \rho e^{i(\psi + \omega_L \tau)} \vb*{g}_n }^2 \left( \ev{\sigma_+ \sigma_-} - \abs{\ev{\sigma_-}}^2 \right) \nonumber \\
        & \quad + 2 \rho \Re \sum_n \vb*{f}_n^\dagger \vb*{g}_n e^{i(\psi + \omega_L \tau)} \left( \ev{\sigma_{+,H}(\tau) \sigma_-} - \ev{\sigma_+ \sigma_-} \right).
\end{align}
We set $e^{i(\psi + \omega_L \tau)} = -1$ for destructive interference and use $\vb*{F}=\vb*{G}$, so the above reduces to
\begin{align}\label{eq_app:Idet}
    I^\mathrm{det} & = (1-\rho)^2 \abs{\vb*{F}}^2 \abs{\ev{\sigma_-}}^2 \nonumber \\
        & \quad + \left( \ev{\sigma_+ \sigma_-} - \abs{\ev{\sigma_-}}^2 \right) \sum_n \abs{\vb*{f}_n - \rho \vb*{g}_n}^2 \nonumber \\
        & \quad + 2 \rho \left( \ev{\sigma_{+,H}(\tau) \sigma_-} - \ev{\sigma_+ \sigma_-} \right) \Re \sum_n \vb*{f}_n^\dagger \vb*{g}_n.
\end{align}

\subsection{Saturation}
In the case of perfect reflectivity $\rho=1$, the first term vanishes and we are left with two remainder terms,
\begin{align}
    R_1 & = C_1(\gamma,\Omega,\Delta) \sum_n \abs{\vb*{f}_n - \vb*{g}_n}^2 , \nonumber \\
    R_2 & = 2 C_2(\tau,\gamma,\Omega,\Delta) \Re \sum_n \vb*{f}_n^\dagger \vb*{g}_n.
\end{align}
The overall function modulating $R_1$ is found in the driven model to be
\begin{align}
\label{eq_app:c_1}
    C_1(\gamma,\Omega,\Delta) = \ev{\sigma_+ \sigma_-} -\abs{\ev{\sigma_-}}^2 & = \frac{\Omega^4}{2(\Omega^2 + 2\Delta^2 + 2\gamma^2)^2}.
\end{align}
For weak driving, $C_1 \approx \Omega^4 / 8 (\Delta^2 + \gamma^2)^2$ -- i.e., this is suppressed to fourth order. Note that this term is analogous to the incoherent far-field intensity found in Eq.~\eqref{eq:farfield_noSLM}.

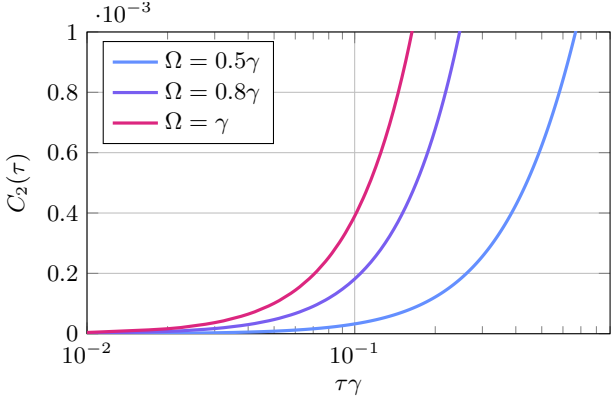
\begin{figure}[t!]
    \centering
    \begin{tikzpicture}
	\begin{semilogxaxis}[	
 width=\textwidth/2.1, height=\axisdefaultheight/1.3, 
            xmin = 0.01, xmax = 0.9,
            ymin = 0, ymax = 0.001,
            xlabel = {$\tau\gamma$},
	    ylabel = {$C_2(\tau)$},
     	legend cell align = {left},
            grid=major,
	      legend pos = north west,]
     
    \addplot[
	    smooth,
	    very thick,
	    color1,] 
     file[skip first] {Figures/Invisibility/Temporal/intensity_1.dat};

     \addplot[
	    smooth,
	    very thick,
	    color2,] 
     file[skip first] {Figures/Invisibility/Temporal/intensity_2.dat};

    \addplot[
	    smooth,
	    very thick,
	    color3,] 
     file[skip first] {Figures/Invisibility/Temporal/intensity_3.dat};

\legend{
	$\Omega=0.5\gamma$, 
	$\Omega=0.8\gamma$,
	$\Omega=\gamma$
}
\end{semilogxaxis}
    \end{tikzpicture}
    
    \caption{The function $C_2$ as defined in Eq.~\eqref{eq:c_2} modulating the mean intensity $I^{\mathrm{det}}$ remaining on the detector, for different driving strengths $\Omega$ and with vanishing detuning. Realistic experimental conditions are $\tau\gamma=0.027$, at which the value of $C_2(\tau)$ is small enough that any leftover field is negligible, independently on the drive.}
    \label{fig:c2}
\end{figure}

\subsection{Temporal coherence}
The second function, which also depends on the time delay, is 
\begin{equation}
\label{eq_app:c_2}
    C_2(\tau,\gamma,\Omega,\Delta) = \ev{\sigma_{+,H}(\tau) \sigma_-} - \ev{\sigma_+ \sigma_-}\; .
\end{equation}
The $\tau$-dependent correlator can be evaluated using the quantum regression theorem as
\begin{align}
\label{eq_app:tau_correlator}
    \ev*{\sigma_{+,H}(\tau) \sigma_-} & = \tr \left[ \sigma_+ e^{\tau \mathcal{L}}(\sigma_- \rho_\infty) \right] \nonumber \\
        & = \tr \left[ \sigma_+ e^{i \omega_L \tau \sigma_z/2} e^{\tau \mathcal{L}_I}(\sigma_- \rho_\infty) e^{-i \omega_L \tau \sigma_z/2} \right] \nonumber \\
        & = e^{i \omega_L \tau} \tr \left[ \sigma_+ e^{\tau \mathcal{L}_I}(\sigma_- \rho_\infty) \right],
\end{align}
where $\mathcal{L}$ and $\mathcal{L}_I$ are the lab-frame and interaction-picture Lindblad generators, according to Eqs.~\eqref{eq_app:master} and \eqref{eq_app:master_interaction} respectively.
These functions are illustrated in Fig.~\ref{fig:c1}.\\

Now suppose that all ions have the same image translated by different amounts -- i.e., $\vb*{f}_n(\vb*{r}) = \vb*{f}(\vb*{r} - \vb*{a}_n)$.
Then the Fourier transform is $\tilde{F}(\vb*{k}) = \tilde{\vb*{f}}(\vb*{k}) \tilde{\vb*{h}}(\vb*{k})$, with $\tilde{\vb*{h}}(\vb*{k}) = \sum_n e^{-i \vb*{a}_n \cdot \vb*{k}}$.
If we further assume the points $\vb*{a}_n$ are symmetrically distributed around the origin, then $\tilde{\vb*{h}}$ is real.
It follows that the SLM phase that works for a single image $\vb*{f}$ also works for the whole image $\vb*{F}$.
However, each position $\vb*{a}_n$ gets inverted, so $\vb*{g}_n(\vb*{r}) = \vb*{f}(\vb*{r} + \vb*{a}_n)$.
In this case, $R_1$ still does not vanish.
Given that the images are spatially non-overlapping, the only term remaining in the sum in $R_2$ is an ion at the origin.

\section{Contribution of the ion motion} \label{sec:app_ionmotion}

In this section, we take into account the motion of the ions, including its dependence on temperature.
We assume that the ions' positions fluctuate in time, writing $\vb*{f}_n = \vb*{f}_n^0 + \delta \vb*{f}_n$, where the mean value of the fluctuating part vanishes, $\ev{\delta \vb*{f}_n} = 0$, and we assume that they are uncorrelated, so $\ev{\delta \vb*{f}_m \delta \vb*{f}_n} = 0$.
The possible effects of this motion are:
\begin{itemize}
    \item Difference in $\vb*{f}_n$ and $\vb*{g}_n$ due to the ion having moved during the round-trip time $\tau$. The oscillation timescale is $\tau_\mathrm{osc} \sim \SI{e-6}{s}$, whereas $\tau \sim \SI{0.1}{m} / c \sim \SI{3e-10}{s} \ll \tau_\mathrm{osc}$. Therefore we can ignore this.
    \item Degraded fidelity of the SLM image compared with the direct image, since the function $m$ is chosen according with the equilibrium-position field $\vb*{F}^0$, i.e., such that $\vb*{G}^0 = \vb*{F}^0$.
\end{itemize}
We focus on the lowest-order contribution of the second effect, which alters the first term in Eq.~\eqref{eqn:i_det1}.
Disregarding the final two remainder terms, we have the average intensity
\begin{align}
    \ev{I^\mathrm{det}} & \approx \abs{\ev{\sigma_-}}^2 \ev{ \abs{\sum_n \left( \vb*{f}_n^0 + \delta \vb*{f}_n - \vb*{g}_n^0 - \delta \vb*{g}_n \right) e^{i\omega_L s_n/c} }^2 } \nonumber \\
        & = \abs{\ev{\sigma_-}}^2 \ev{ \abs{\sum_n \left( \delta \vb*{f}_n - \delta \vb*{g}_n \right) e^{i\omega_L s_n/c} }^2 } \nonumber \\
        & = \abs{\ev{\sigma_-}}^2 \sum_n \ev{ \abs{\delta \vb*{f}_n - \delta \vb*{g}_n}^2 },
\end{align}
having used the assumption that the ions are uncorrelated in their motion.
Since the fluctuations in the images derive from position fluctuations, we have
\begin{align}
    \vb*{f}_n(\vb*{r}) & = \vb*{f}_n^0(\vb*{r} + \delta \vb*{r}_n) e^{i k \delta z_n} \nonumber \\
        & \approx \vb*{f}_n^0(\vb*{r}) + (\delta \vb*{r}_n \cdot \vb*{\nabla}) \vb*{f}_n^0(\vb*{r}) + i k \delta z_n \vb*{f}_n^0(\vb*{r}) ,
\end{align}
where we assume the in-plane fluctuations $\delta \vb*{r}_n$ are small compared with the image scale and the $z$-axis fluctuations $\delta z_n$ are small compared with the wavelength.
Note that, since $\vb*{g}_n$ is shifted in the opposite direction and the relevant path length is changed by $-\delta z_n$, we have $\delta \vb*{g}_n \approx - (\delta \vb*{r}_n \cdot \vb*{\nabla}) \vb*{g}_n^0 - ik\delta z_n \vb*{g}_n^0$.
We take the fluctuations to satisfy $\ev{\delta r_{n,i}} = 0$ and a coordinate system such that $\ev{\delta r_{n,i} \delta r_{n,j}} = \delta_{ij} \sigma_i^2$ (also assuming the $z$-axis motion is uncorrelated with the in-plane motion), so
\begin{align}
    \ev{\abs{\delta \vb*{f}_n - \delta \vb*{g}_n}^2} & = \ev{ \abs{ (\delta\vb*{r}_n \cdot \vb*{\nabla} + ik \delta z_n) (\vb*{f}_n^0 + \vb*{g}_n^0) }^2 } \nonumber \\
        & = \sum_{i,j=x,y} \ev{\delta r_{n,i} \delta r_{n,j}} (\partial_i [\vb*{f}_n^0+\vb*{g}_n^0])^\dagger \nonumber \\
        & \qquad \qquad \qquad \qquad \cdot (\partial_j [\vb*{f}_n^0 + \vb*{g}_n^0]) \nonumber \\
        & \qquad + k^2\ev{\delta z_n^2} \abs{\vb*{f}_n^0 + \vb*{g}_n^0}^2 \nonumber \\
        & = \sum_{i=x,y} \sigma_i^2 \abs{ \partial_i [\vb*{f}_n^0 + \vb*{g}_n^0] }^2 \nonumber \\
        & \qquad +  k^2 \sigma_z^2 \abs{\vb*{f}_n^0 + \vb*{g}_n^0}^2.
\end{align}

We now take the images as $\vb*{f}_n^0(\vb*{r}) = \vb*{f}(\vb*{r}-\vb*{a}_n)$, $\vb*{g}_n^0(\vb*{r}) = \vb*{f}(\vb*{r} + \vb*{a}_n)$, so
\begin{align}
    \ev{\abs{\delta\vb*{f}_n - \delta\vb*{g}_n}^2} \approx \sum_i \sigma_i^2 \abs{ \partial_i[\vb*{f}(\vb*{r}-\vb*{a}_n) + \vb*{f}(\vb*{r}+\vb*{a}_n)]}^2 \nonumber \\
         \approx 
        \begin{cases}
            \sum_{i=x,y} 4 \sigma_i^2 \abs{\partial_i[\vb*{f}(\vb*{r})]}^2 + 4 k^2 \sigma_z^2 \abs{\vb*{f}_n^0}^2, & \vb*{a}_n = \vb*{0} \\
            \sum_{i=x,y} \sigma_i^2 \left[ \abs{\partial_i \vb*{f}(\vb*{r}-\vb*{a}_n)}^2 + \abs{\partial_i \vb*{f}(\vb*{r}+\vb*{a}_n)}^2 \right] & \\
            \quad + k^2 \sigma_z^2 \left[ \abs{\vb*{f}(\vb*{r}-\vb*{a}_n)}^2 + \abs{\vb*{f}(\vb*{r}+\vb*{a}_n)}^2 \right] & \vb*{a}_n \neq \vb*{0},
        \end{cases}
\end{align}
where we assume that the displaced images are non-overlapping.\\

As a simple model, we take a scalar image with an isotropic Gaussian profile $f(\vb*{r}) = \kappa e^{-\abs{\vb*{r}}^2 / 4 s^2}$.
This represents the amplitude arriving at the detector directly from the ion, taking into account a point spread function from the imaging optics.
(We neglect any additional aperture for the round-trip amplitude via the SLM.)
Then
\begin{align} \label{eqn:residual_image}
    &\ev{\abs{\delta f_n - \delta g_n}^2} \approx  \nonumber \\
    &    \begin{cases}
            \kappa^2 \left[ \sum_i  \left(\frac{\sigma_i r_i}{s^2}\right)^2 + 4k^2 \sigma_z^2 \right] e^{-\abs{\vb*{r}}^2/2s^2} , & \vb*{a}_n = \vb*{0} \\
            \kappa^2 \left[ \sum_i \left(\frac{\sigma_i (r_i - a_{n,i})}{2s^2}\right)^2  + k^2 \sigma_z^2 \right] e^{-\abs{\vb*{r}-\vb*{a}_n}^2/2s^2} \\
             + \kappa^2  \left[ \sum_i \left(\frac{\sigma_i (r_i + a_{n,i})}{2s^2}\right)^2 + k^2 \sigma_z^2   \right]e^{-\abs{\vb*{r}+\vb*{a}_n}^2/2s^2} , & \vb*{a}_n \neq \vb*{0}.
        \end{cases}
\end{align}
We now assume an isotropic potential of uniform scale $\sigma$.
Compared with the intensity of the bare image, the above correction is approximately suppressed by a factor
\begin{align} \label{eq_app:motion_factor}
    C_3 := \left(\frac{\sigma}{2s}\right)^2 + (k \sigma)^2.
\end{align}
For the width $s$, we take the Abbe diffraction limit, which for wavelength $\lambda = \SI{493}{nm}$ and numerical aperture $\mathrm{NA}=0.6$ evaluates to $s = \lambda / 2 \mathrm{NA} \approx \SI{411}{nm}$.
Note that $2s k = 2\pi / \mathrm{NA} \approx 10.5$, so the second term in Eq.~\eqref{eq_app:motion_factor} (arising from motion along the optical axis) is dominant.

The standard deviation in position of a quantum oscillator of mass $m$ in a harmonic trap of frequency $\omega$ at temperature $T$ is 
\begin{align}
    \sigma = \sqrt{\frac{(2\bar{n} + 1)\hbar}{2\omega m}},
\end{align}
which depends on the mean phonon number
\begin{align}
    \bar{n} = \frac{1}{e^{\hbar \omega / k_B T}-1} .
\end{align}
For a Barium ion and $\omega = 2\pi \times \SI{1}{\mega \hertz}$, for the ground state we have $\sigma \approx \SI{6.1}{nm}$ and so $(\sigma / s)^2 \approx \num{2e-4}$.
At thermal occupations of $\bar{n} = 10$ (the Doppler cooling limit) and $200$, this factor evaluates to $(\sigma / s)^2 \approx \num{5e-3}$ and $\num{9e-2}$ respectively.

\bibliography{main}

\end{document}